\documentstyle[sprocl,epsfig]{article}

\def\Journal#1#2#3#4{{#1} {\bf #2}, #3 (#4)}

\def\PLB{{\em Phys. Lett.}  B}

\def\be{\begin{equation}}
\def\ee{\end{equation}}
\def\bea{\begin{eqnarray}}
\def\eea{\end{eqnarray}}
\def\ifmath#1{\relax\ifmmode #1\else $#1$\fi}%

\def\rW{\ifmath{{\mathrm{W}}}}
\def\rZ{\ifmath{{\mathrm{Z}}}}
\def\fm{\ifmath{{\mathrm{fm}}}}
\def\noBE{\ifmath{{\mathrm{noBE}}}}
\def\MC{\ifmath{{\mathrm{MC}}}}
\def\data{\ifmath{{\mathrm{data}}}}
\def\diff{\ifmath{{\mathrm{diff}}}}
\def\same{\ifmath{{\mathrm{same}}}}
\def\mixed{\ifmath{{\mathrm{mixed}}}}
\def\hadr{\ifmath{{\mathrm{hadr}}}}

\begin{document}

\title{INTER-W BOSE-EINSTEIN CORRELATIONS ... OR NOT?}

\author{\v S.TODOROVA-NOV\'A}

\address{CERN, CH-1211 Geneva, Switzerland\\E-mail:todorovova@cern.ch}

\maketitle\abstracts{A critical summary is given of the present status 
of the study of Bose-Einstein Correlations in W-pair production at LEP II.
In particular, the evidence is reviewed for or against the existence of 
Bose-Einstein correlations between pions originating both from
a different of the two W's. If present, such an inter-W interference would 
not only form a potential bias in the determination of the W
mass, but also would provide a laboratory to measure the space-time 
development of the overlap. If absent, this would drastically
   change the conventional (Hanbury Brown and Twiss) picture of pion 
interferometry in high energy physics. 
}

\section{Introduction}

    Correlations between pairs of identical particles (or, in the simplified
  experimental approach, pairs of like-sign particles) within a single
  hadronic system are a well known phenomenon, however the understanding
  of this effect is far from complete. Most often, it is considered
  to be an equivalent of the Hanbury Brown and Twiss effect in astronomy,
  reflecting the interference of identical bosons  
  emitted {\it incoherently} from their source.

    An alternative model, proposed by B.Andersson
  and collaborators,\cite{lund} takes into account the full process
  of particle production in the fragmentation of the Lund string. 
  The correlations appear as a {\it coherent} effect in the hadronization
  process, and they are fully predicted for a given set of final
  particles (ordered along the string). 

    We are, therefore, in the situation that the experimentally observed
  correlations can be interpreted in two rather different ways:
  in the `incoherent' approach, the shape of the correlation function
  reflects the shape of the source, and can be derived from the knowledge
  of the space-time density of the final particles regardless of
  the way they were produced. In the `coherent' picture, the correlations
  stem directly from the string area decay law, and depend on the
  history of the string breaking. 

    For a simple hadronic system like $q\bar{q}$ from a Z$^0$ decay, it
  may be impossible to decide between the two possibilities, since the 
  incoherent approach leaves the freedom of choice of the input
  particle density, which can be adjusted to reproduce the observed data 
  (it should be noted, however, that 
  the straightforward implementation
  of the `incoherent' formalism fails to describe the Z$^0$ data
~\cite{heinz}). 
  
    The situation is different in the study of two (partially)
  overlapping hadron\-ic systems, see ~\cite{dewolf}. In the incoherent
  scenario, the difference between correlations
  within a single hadronic system, and correlations between the two systems, 
  should depend only on the overlap of the two systems (sources).
  In the coherent scenario,
  however, the correlations between the two systems may 
  not exist at all, even for overlapping sources (as long as there is no
  interaction -color flow- between them).

 \section{Measurements} 

     In the light of the discussion above, the experimental
  measurement of correlations between two independent hadronic systems
  is of utmost interest, and LEP2 provides a unique laboratory
  for such a measurement in the study of the decay of a pair
  of $\rW^+\rW^-$ (resp.$\rZ^0\rZ^0$) bosons. The life-time of these bosons
  is much shorter than the typical hadronization scale,
  and they decay on top of each other, overlapping (partially)
  in momentum space.    

     The measurement of the size of the inter-W(Z) correlations
   can be done in different ways, and it is complicated by several factors,
   namely: 
\begin{itemize}
\item the modeling of the effect is poor (none of the models discussed
  in the Introduction is fully implemented in MC generators); the most
  widely used model (LUBOEI/PYBOEI in Jetset ~\cite{jetset}) consists
  in a simple reshuffling of momenta of final particles,
  leading to an artificial momentum transfer 
\item the effect is defined with respect to a reference (uncorrelated)
   sample, which is arbitrary to a large extent, and different for
   each collaboration/measurement
\item detector effects can be important, and not easy to correct for
   because of model dependence of the correction factors for 2-particle 
spectra.
\end{itemize}
   The experimental methods used by the LEP collaborations
  for this measurement can be roughly classified according to the level
  of their model dependence.
  
  The `model dependent' methods consist in tuning of a particular model
  at the Z$^0$/single-W decay and in comparison of the prediction
  of the model with real WW/ZZ events. Such a method was used by 
ALEPH.\cite{aleph1} The correlations
  between the like-sign particles (after rejection of identified electrons
  and muons) are defined with
  respect to the unlike-sign particles sample, and the double ratio with
  the Monte-Carlo sample is used
  to remove the effect of resonances as well as part of detector effects:

  \begin{equation}
   R^*(Q)=(\frac{N^{++,--}}{N^{+-}})^{\data} /
            (\frac{N^{++,--}}{N^{+-}})^{\MC}_{\noBE} 
   \label{dratio}
  \end{equation}

\begin{figure}[t]
 \mbox{\epsfig{file=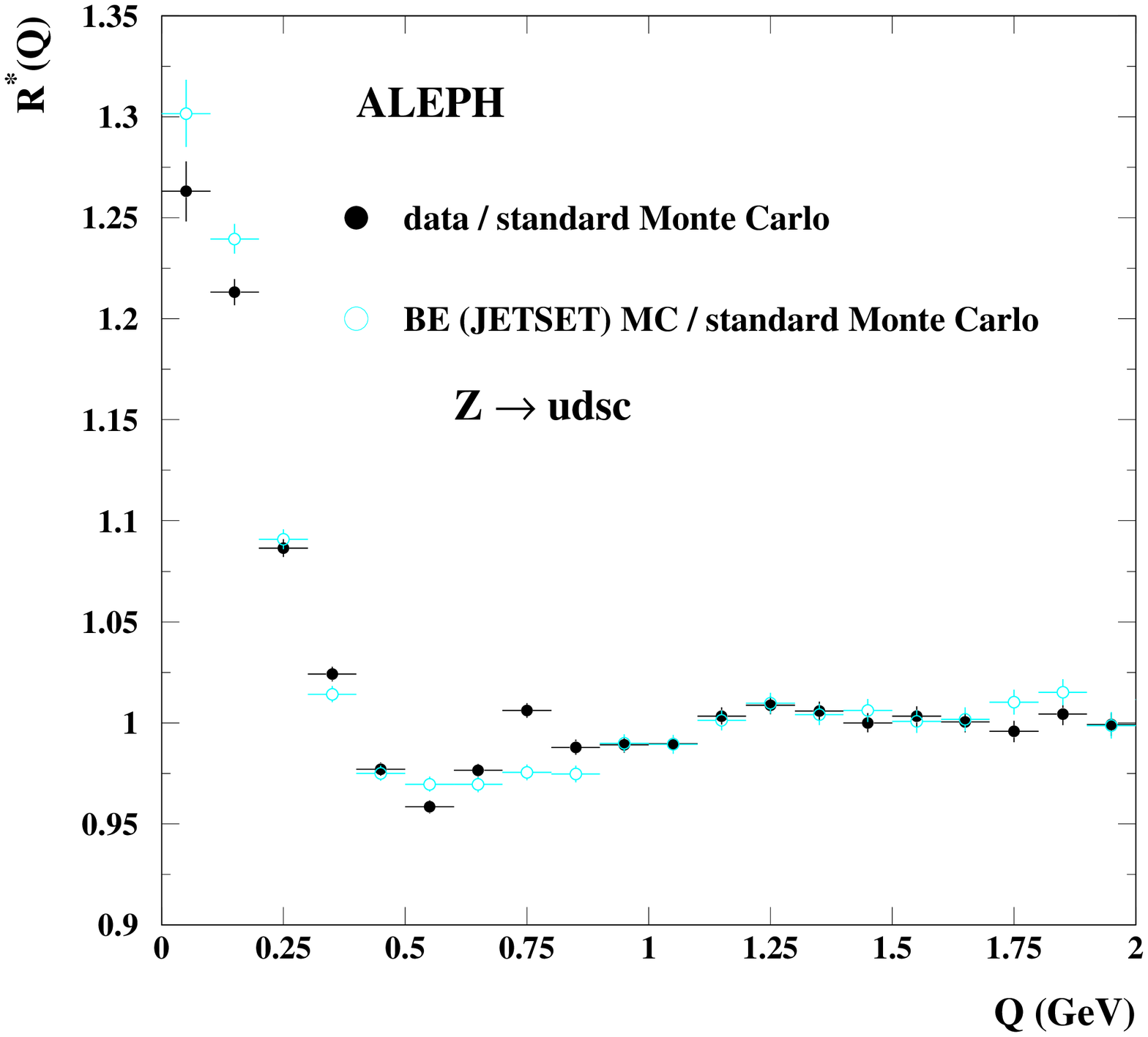,height=6.cm,width=6.cm}%
       \epsfig{file=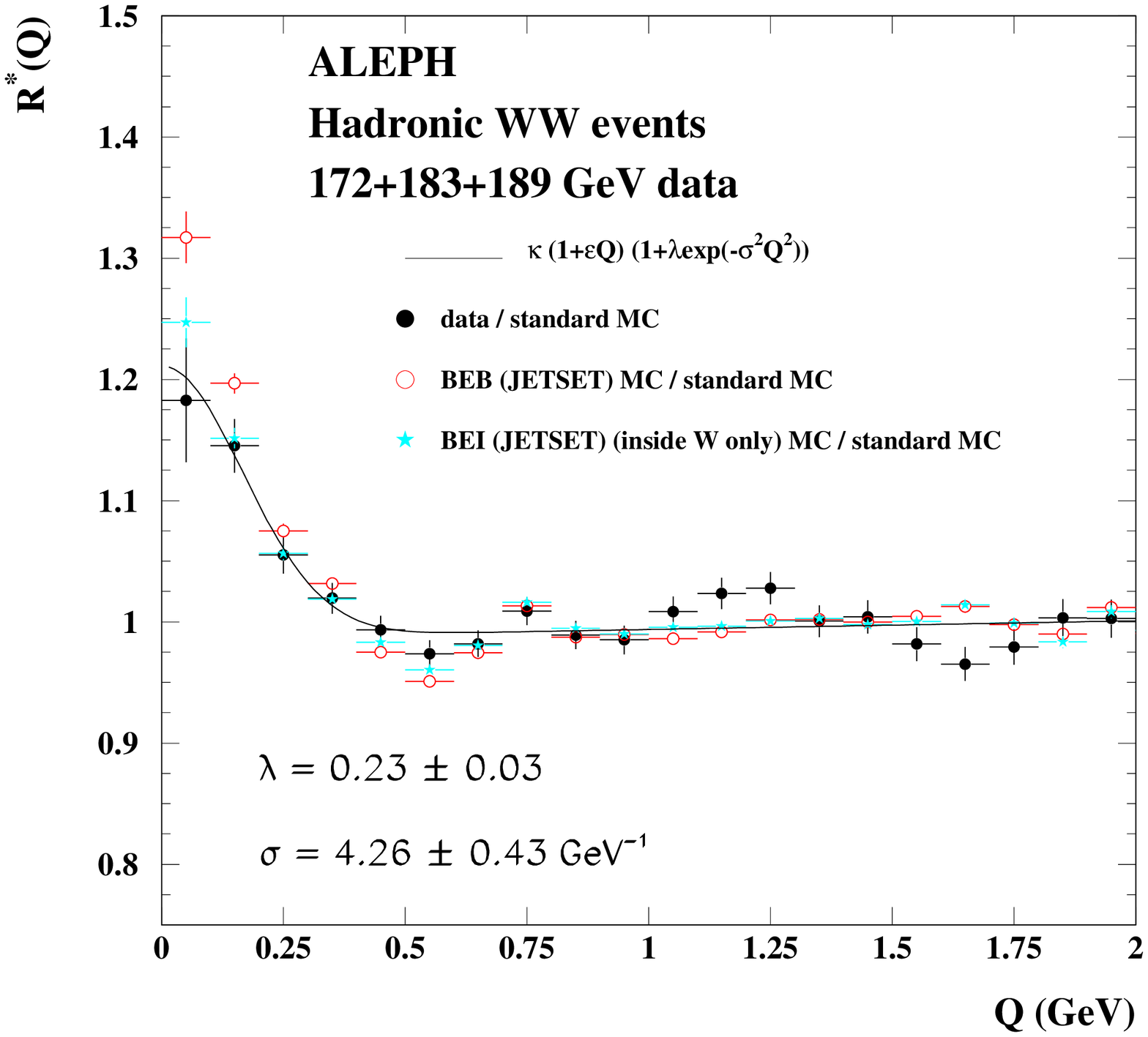,height=6.cm,width=6.cm}}
\caption{The double ratios $R^*$ measured in the Z$^0$ and in the
  fully hadronic W$^+$W$^-$ sample, compared to the prediction of the model
  tuned at Z$^0$ (PYBOEI BE$_3$).
 \label{fig:aleph}}
\end{figure}

  The tuning of the PYBOEI routine is performed on the Z$^0$ sample
  enriched in light flavours and checked in the semileptonic W events.
  The residual discrepancies between data and simulation are corrected
  bin per bin and correction is applied on MC predictions for fully hadronic
  events, in the scenario with and without correlation between the
  two boson systems (Fig.~\ref{fig:aleph}). The $q\bar{q}$ background
  is included in the MC prediction.   The data disfavour the presence
  of inter-W/Z correlations by 2.7 $\sigma$ (for this particular tuning 
of the model).

   Apart from a strong model dependence built into this measurement, one
 may worry also about the fact that due to the use of double ratios, the
 model may be actually quite far from the data in a direct comparison
 (not reproducing the Q distribution, itself).

\begin{figure}[bth]
\begin{tabular}{ll}
\begin{minipage}{6.cm}
\mbox{\epsfig{file=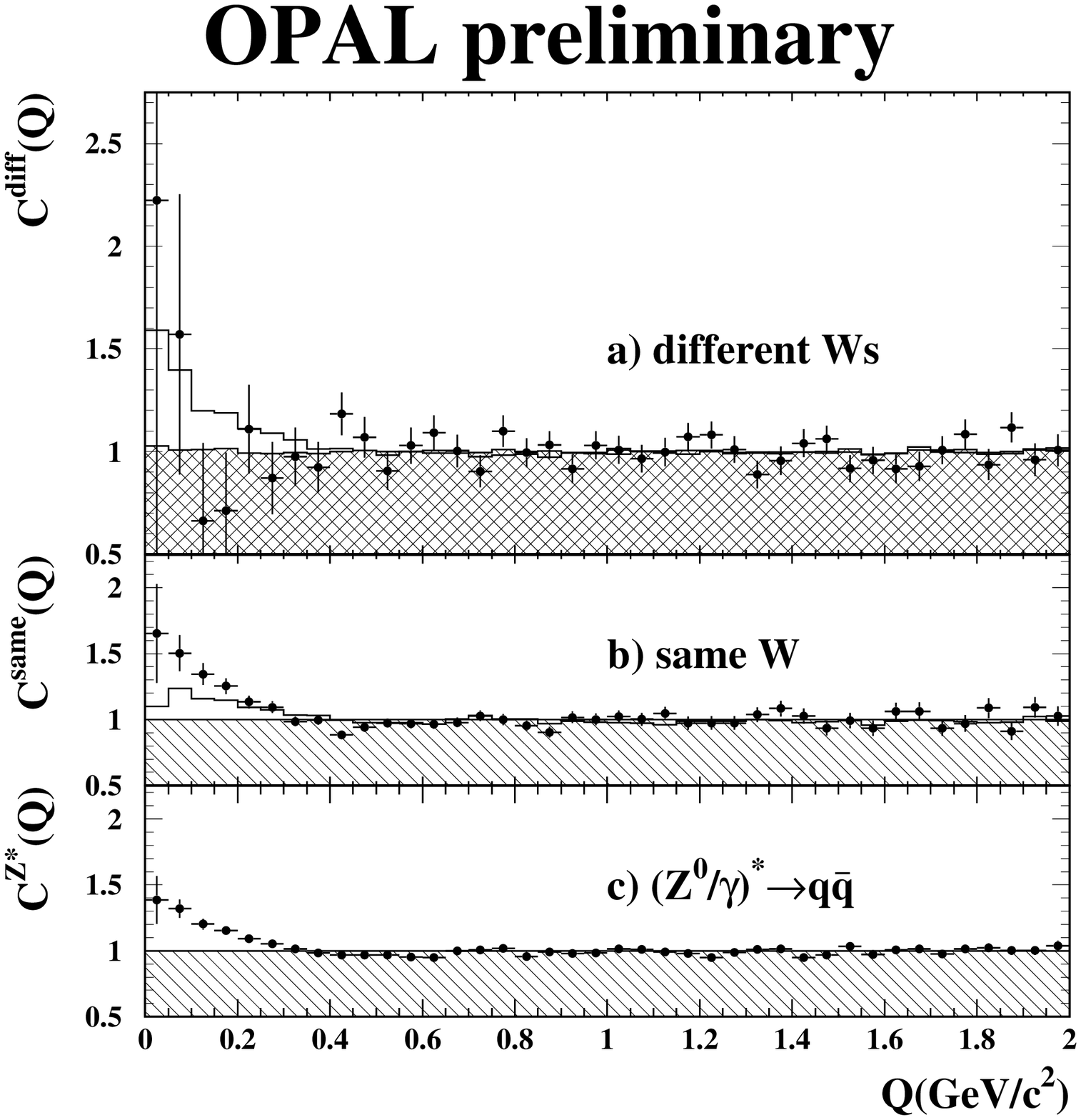,height=7.cm,width=5.2cm}}
\end{minipage}  & 
\begin{minipage}{6.cm}
   Results of the simultaneous fit:\\

  $\lambda^{\diff}=0.05\pm0.67\pm0.35$  \\
  $R^{\diff}=1.51\pm0.05\pm0.09 / \fm $ \\

  $\lambda^{\same}=0.69\pm0.12\pm0.06$ \\
  $ R^{\same}=1.07\pm0.07\pm0.12 / \fm $ \\

  $\lambda^{\rZ^*}=0.43\pm0.06\pm0.0$ \\
  $ R^{\rZ^*}=1.01\pm0.08\pm0.14 / \fm $ \\

(data at 172, 183 and 189 GeV) \\

\end{minipage}
\end{tabular}
\caption{Correlation function for the unfolded classes. The data points show
 the experimental distributions. The open histogram shows a) the result
 of the simulation including inter-W correlations, b) the result of simulation
 including correlations within a single W. The cross-hatched histogram in a)
 shows result of simulation with correlation only within a single W, while
 the hatched histograms in b) and c) correspond to a simulation without
 any BE correlations.}      
\label{fig:opal}
\end{figure}

  A different method with lower model dependence was used by OPAL.\cite{opal}
 It is based on
 a simultaneous fit of the correlation functions in Z$^0/\gamma^*$,
  $q\bar{q}l\nu$, and $q\bar{q}q\bar{q}$ events. The Z$^0$/WW content
  in these three samples is parametrized according to the selection
  efficiency obtained with the MC simulation. A correlation function,
  defined as the double ratio (\ref{dratio}), is extracted
  for pairs of particles coming from $\rZ^0/\gamma$, single W, and from 
  different W's in a simultaneous fit. The result is shown 
in Fig.~\ref{fig:opal}.   
 Unfortunately, due to the large uncertainties, the method is far from being
 sensitive to the effect of inter-boson correlations ($\lambda^{\diff}$),
 and no conclusions about the presence of these correlations can
 be made.  
 
  In the measurement published by L3,\cite{l3} the model dependence
  is largely removed due to the use of a reference sample constructed by
  mixing of the hadronic parts of the semileptonic W$^+$W$^-$ events.
  Such a direct `data-to-data' comparison \cite{ckw} is experimentally robust
  and does not require a direct use of MC simulation (except
  for checking of the mixing method). The only residual model
  dependence is related to the subtraction of background events.  

\begin{figure}[h]
 \mbox{\epsfig{file=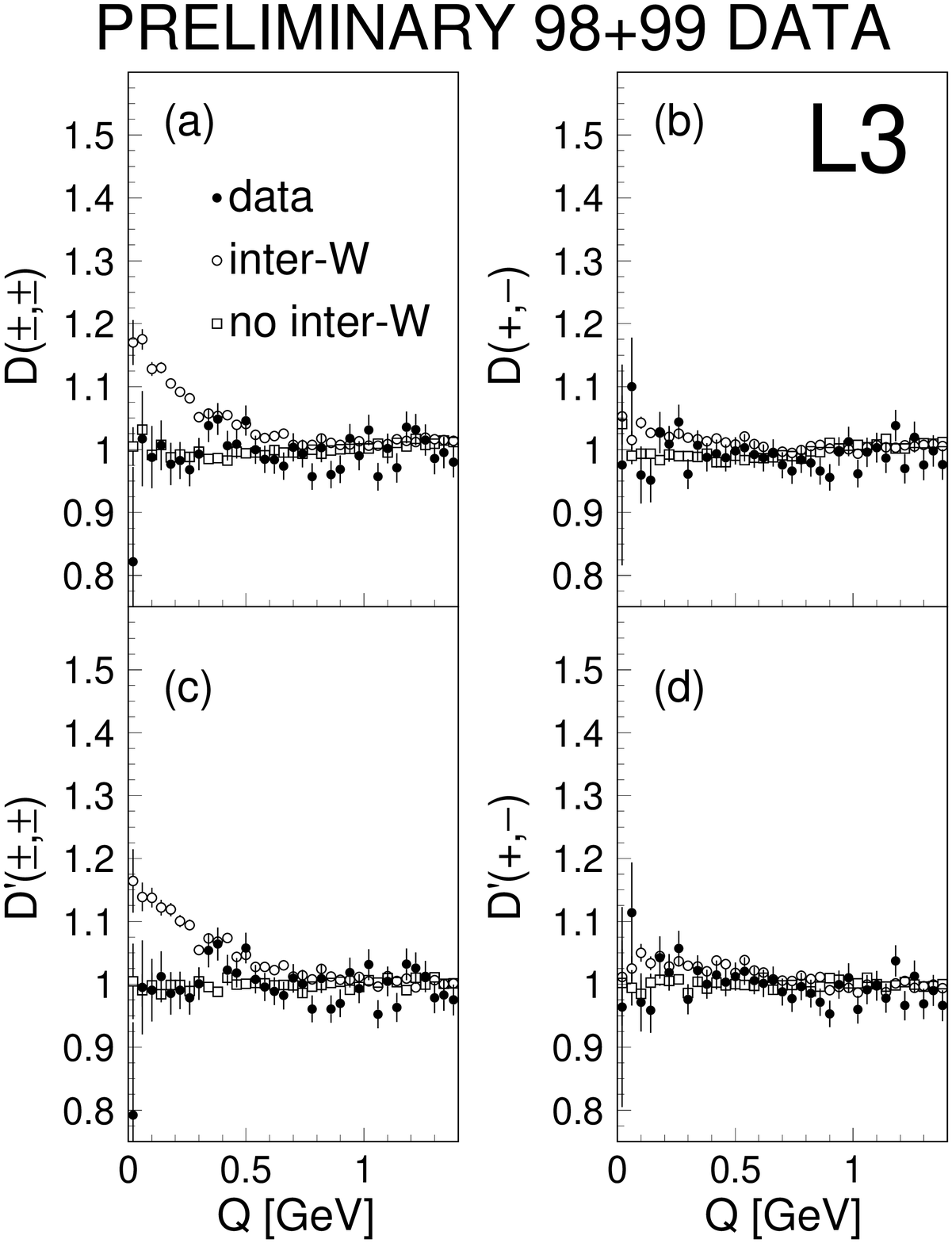,height=10.cm,width=11.cm}}
\begin{tabular}{ll}
\begin{minipage}{6.cm}
\mbox{\hspace{0.5cm}\epsfig{file=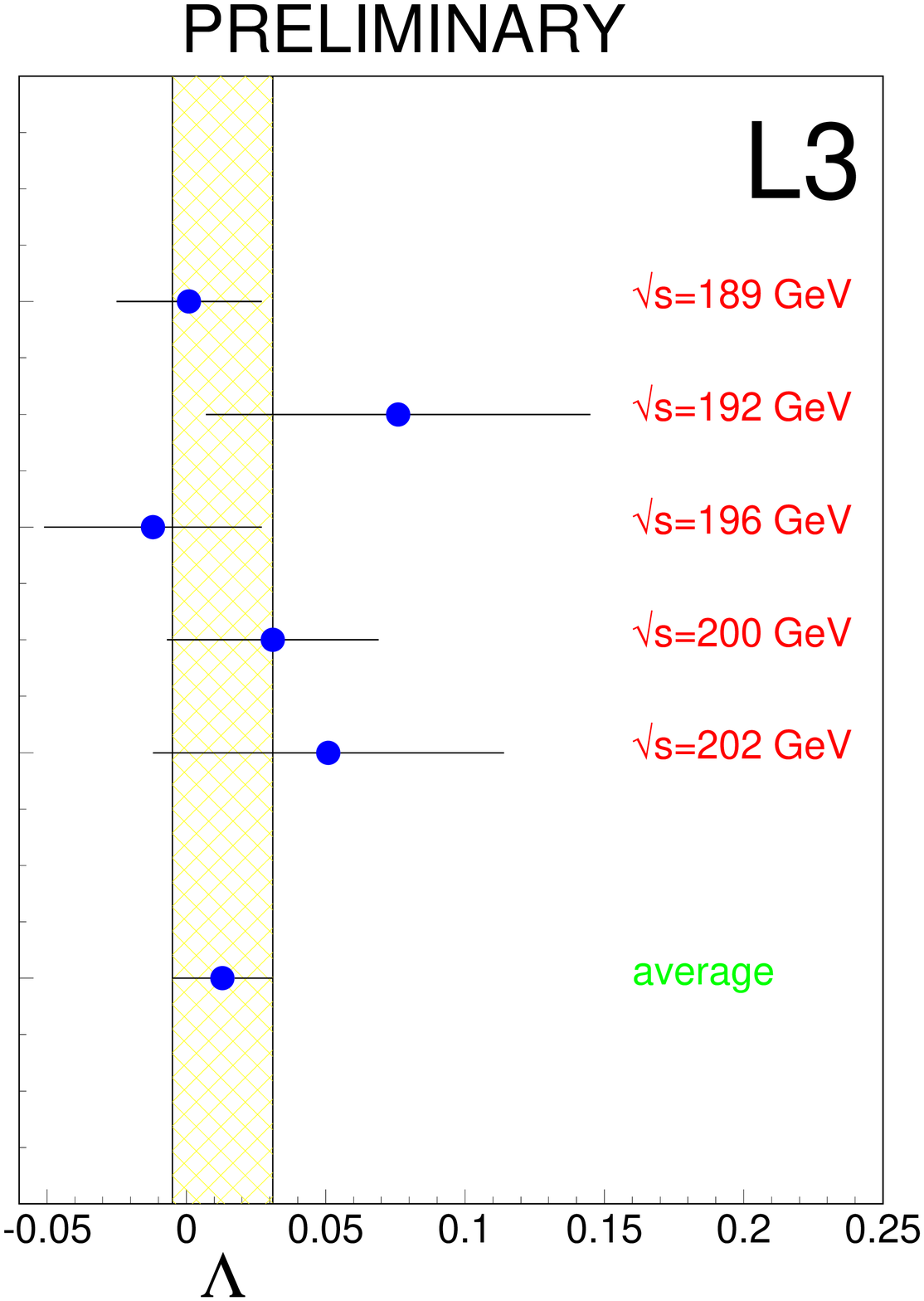,width=5.cm}}
\end{minipage} 
& \begin{minipage}{6.cm}
  Fit of the shape of D'($\pm,\pm$)  \\
  with the function \\
 $(1+\epsilon Q)(1+\Lambda \exp{-k^2Q^2)}$  \\
  gives the following result \\
  (combined 98+99 data): \\

  $\Lambda=0.013\pm0.018\pm0.015$  

\end{minipage}
\end{tabular}
\caption{The results obtained by L3 using the mixing method, for like-sign
 and unlike-sign pairs in WW events, compared to the prediction of the
 tuned model (PYBOEI BE$_{32}$). } 
\label{fig:l3}
\end{figure}

   Fig.~\ref{fig:l3} shows the ratios of 2-particle densities obtained
  from the fully hadronic W$^+$W$^-$ sample and from the mixed sample:
\begin{equation}
    D(Q)=\rho_2^{\hadr.\rW\rW}(Q)/\rho_2^{\mixed \ \rW\rW}(Q) \\
\end{equation}
   and the  double ratio:
\begin{equation}
    D'(Q)= D^{\data}(Q)/D^{\MC,\noBE}(Q).
\end{equation} 

    There is no evidence for the existence of inter-W correlations in L3
  results.

   Since the model dependent measurements have only very limited
  impact, the advantages of the method used by L3 seem to be acknowledged
 by the other LEP experiments which are preparing similar measurements.
 Recently, ALEPH released preliminary results using the mixed reference
 sample.\cite{aleph2} 
   
\clearpage

 The results cannot be
 quantified because of missing systematic errors. However the measurement
 seems to disfavour the presence of correlations, in agreement with
 L3 and with aforementioned ALEPH results.  
   
  The DELPHI experiment is the only experiment which reported
 a hint for presence of inter-W correlations.\cite{delphi}
 However, some problems were found
 in the analysis, and the observation of the effect was not confirmed
 by an independent analysis. The discrepancies are yet to be clarified,
 and for the moment the results are uncertain.    

\section{Combination of LEP results}
  Due to the large variety of analyses performed so far, it is impossible
  to combine the results in a single number. Qualitatively, there is no
  confirmed observation of inter-W correlations. In the future, it seems
  probable that LEP collaborations will converge to a single method
  \`a la L3, which provides the most direct, 
  and less biased, access to the
  inter-W correlations. Still, even using similar analysis method, there
  are quite a few problems to be solved before the combination can be done.
  Ideally, if the data would be corrected for the detector effects, a direct
  combination of measured distributions would be possible, providing
  at the same time a cross-check of compatibility of individual measurements.
  If for some reasons the data cannot be corrected for detector effects,
  one has to use a model for comparison between experiments. A unique
  choice of the model for such a comparison is a non-trivial task; 
  the tuning of the
  model parameters vary significantly between experiments, and it is
  intertwined with the tuning of parameters of fragmentation models.
  In any case, a rather close collaboration between experiments is required
  in order to reach a combined LEP result.

\end{document}

%%%%%%%%%%%%%%%%%%%%%%%%%%%%%%%%%%%%%%%%%%%%%%%%%%%%%%%%%%%%%%%%%%%%%%%%%%
%% End of sprocl.tex  
%%%%%%%%%%%%%%%%%%%%%%%%%%%%%%%%%%%%%%%%%%%%%%%%%%%%%%%%%%%%%%%%%%%%%%%%%%